%% file: Compton_PRL.tex
\documentstyle[eqsecnum,aps,epsfig,twocolumn]{revtex}

\begin{document}




\newcommand{\dis}{\displaystyle}
\newcommand{\nn}{\nonumber}
\newcommand{\aexp}{A_{exp}}
\newcommand{\araw}{A_{raw}}
\newcommand{\ath}{A_{{th}}}
\newcommand{\ac}{A_{{c}}}
\newcommand{\af}{A_{{F}}}
\newcommand{\pe}{P_{{e}}}
\newcommand{\pg}{P_{\gamma}}
\newcommand{\np}{N^{+}}
\newcommand{\nm}{N^{-}}
\newcommand{\kp}{k^{\prime}}
\newcommand{\krs}{k^{\prime s}_{r}}
\newcommand{\sigrs}{\sigma^{s}_{r}}
\newcommand{\sigc}{\sigma_{c}}
\newcommand{\kmax}{k^{\prime}_{max}}
\newcommand{\wer}{\omega_R}
\newcommand{\wel}{\omega_L}




\draft

\preprint{HEP/123-qed}
\title{
First electron beam polarization measurements with a Compton 
polarimeter at Jefferson Laboratory
}

\author{
M. Baylac,
E. Burtin,
C. Cavata,
S. Escoffier,
B. Frois,
D. Lhuillier,
F. Marie,
J. Martino,
D. Neyret,
T. Pussieux\cite{corauthor}
}

\address{
CEA Saclay, DSM/DAPNIA/SPhN, F-91191 Gif-sur-Yvette Cedex, France, 
http://www.jlab.org/compton
}

\author{
P.Y. Bertin
}

\address{
Université Blaise Pascal et CNRS/IN2P3
LPC, 63 177 Aubi\`ere Cedex, France
}

\author{
C.W. de Jager,
J. Mitchell
}

\address{
Jefferson Lab, 
12000 Jefferson Avenue,
Newport News, VA 23606, USA
}

\date{\today}

\maketitle

\begin{abstract}
A Compton polarimeter has been installed in Hall A at Jefferson 
Laboratory. This letter
reports on the first electron beam polarization measurements 
performed during the HAPPEX 
experiment at an electron energy of 3.3 GeV and an average current of 
40 $\mu$A.

The heart of this device is a Fabry-Perot cavity 
which increased the luminosity for  Compton scattering in the interaction 
region so much that a $1.4\%$ statistical accuracy could be obtained 
within one hour, with a $3.3\%$ total error.  
\end{abstract}


\narrowtext

\section{Introduction}

\label{sec:level1}

Nuclear physics experiments using a polarized electron beam
require an accurate knowledge of the beam polarization.
However, polarization measurements often account for the main 
systematic uncertainty for such experiments.  
The Continuous Electron Beam Accelerator Facility (CEBAF), located at 
Jefferson Laboratory (JLab), is equipped with 
three types of beam polarimeters. A Mott polarimeter, limited to 
low energy (few MeV), is available at the injector, while M\o ller 
polarimetry
is used at low currents (few $\mu$A) in all three experimental halls.
The main limitation of both techniques is their invasive character due to the 
solid target (Au and Fe, respectively) used, preventing beam polarization measurements
simultaneous to running an experiment. Moreover, neither of these
instruments is capable of operating in the energy and/or current regime 
delivered at JLab.
In contrast, Compton polarimetry, which uses elastic scattering of electrons off photons,
is a non-invasive technique.
Indeed, the interaction of the electron beam with the "photon target" 
does not 
affect the properties of the beam so that the beam polarization can be 
measured upstream 
of the experimental target simultaneously with running the experiment. 
The applicability of Compton polarimetry has been demonstrated at 
high-current ($>$ 20 mA) storage rings 
(NIKHEF \cite{Nikhef}, HERA \cite{Barber}) and at high-energy ($>$ 50 
GeV) colliders
(SLAC \cite{Woods}, LEP \cite{LEP}). However, at the CEBAF 
beam conditions (energy of several GeV, current of up to 100 $\mu$A) 
the Compton cross section asymmetry is reduced to a few percent, 
which makes Compton polarimetry extremely challenging.

A pioneering technique was selected to increase the Compton interaction 
rate such 
that a 1\% statistical error measurement could be achieved within an hour: 
an optical Fabry-Perot type cavity fed by an infrared photon source. 
The Compton luminosity has been enhanced further by minimizing the crossing 
angle between the electron and the photon beams, 
which required setting the cavity mirrors only 5 mm away from the electron beam. 
This unique configuration allows measurements of the electron beam 
polarization 
at  moderate beam energies ($>$ 1 GeV) and currents ($>$ 10 $\mu$A).  
The Fabry-Perot cavity \cite{jorda}, comprising two identical high-
reflectivity mirrors, amplifies the photon density at the Compton 
Interaction Point (CIP) with a nominal gain of 7000. 
The initial laser beam (NdYAG, 300 mW, 1064 nm) is locked on the 
resonant frequency of the cavity through an electronic feedback loop \cite{nico}.
The optical cavity, enclosed in the beam pipe, is located at the 
center of a magnetic chicane 
(16 m long), consisting of 4 identical dipoles. The chicane separates the 
scattered electrons and photons
and redirects the primary electron beam unchanged in polarization 
and direction to the Hall A beam line. 
The backscattered photons are detected in a matrix of 25 
PbWO$_4$ crystals 
\cite{neyret}, which are read out by PMT's. The PMT output signal is amplified and 
integrated before being sent to ADC's.
This article reports on the first polarization measurements performed 
during the second part of the data taking of HAPPEX experiment \cite{aniol} 
(July 1999) with the 
Hall A Compton polarimeter at an electron energy of 3.3 GeV and an 
average current of 40 $\mu$A.

\section{Towards a polarization measurement}

\label{sec:toward}

For the HAPPEX experiment CEBAF produced a longitudinally 
polarized electron beam
with its helicity flipped at 30 Hz. This reversal induces 
an asymmetry, $\aexp = \frac{\np-\nm}{\np+\nm}$, 
in the Compton scattering events $N^{\pm}$ detected at opposite helicity.
In the following, the events are defined as count rates normalized 
to the electron beam intensity within the polarization window.
The electron beam polarization is extracted from this 
asymmetry via \cite{prescott}

\begin{equation} 
\dis 
\pe = \frac{\aexp}{\pg \ath},
\label{eq:def}
\end{equation}

\noindent

where $\pg$ denotes the polarization of the photon beam 
and $\ath$ the analyzing power.
The measured raw asymmetry $\araw$   
has to be corrected for 
dilution due to the background-over-signal ratio $\frac{B}{S}$, 
for the background asymmetry $A_B$ and for any helicity-correlated 
luminosity asymmetries $A_F$, so that $\aexp$ can be 
written to first order as

\begin{equation} 
\dis 
\aexp =  \left( 1 + \frac{B}{S} \right) \araw - \frac{B}{S} A_B 
+ A_F.
\label{eq:aexp}
\end{equation}

The polarization of the photon beam can be reversed with a
rotatable quarter-wave plate, allowing asymmetry measurements  
for both photon states, $\araw^{(R,L)}$.
The average asymmetry is calculated as     

\begin{equation}
\dis
\aexp  =  \frac{\wer \araw^R - \wel \araw^L}{\wer + \wel},
\label{eq:asyRL}
\end{equation}

\noindent

where $\omega_{R,L}$ denote the statistical weights of the 
raw asymmetry for each photon beam polarization.
Assuming that the beam parameters remain constant over the 
polarization reversal and that
$\wer \simeq \wel$, false asymmetries cancel out such that

\begin{eqnarray}
\dis
\aexp & \simeq & \frac{\araw^R - \araw^L}{2}  (1 + \frac{B}{S}).  
\label{eq:adil}
\end{eqnarray}

\section{Data taking and event preselection}

\label{sec:data}

Before starting data taking, the electron beam orbit 
and its focus in the central part of the chicane have to be optimized to reduce the 
background rate observed in the photon 
detector which was generated by a halo of the electron beam scraping 
on the Fabry-Perot 
mirror ports \cite{baylac}. The vertical position of the electron beam 
was adjusted until the overlap between the electron and 
the photon beam was maximized at the CIP. This was achieved by 
measuring the count rate in the photon detector while adjusting 
the field of the chicane dipoles. 
Because of a slow drift of the electron beam, its position
had to be readjusted every couple of hours to minimize systematic effects (cf 
section \ref{sec:afalse}).   
Polarimetry data were accumulated continuously while running the 
HAPPEX experiment. 
Each one hour run consisted of five pairs 
of signal - with the cavity resonant (cavity 
ON) - and background - when the cavity was intentionally unlocked (cavity OFF), 
approximatively one third of 
the time - runs. Signal and background events were recorded for
alternating values of the photon beam 
polarization inside the cavity.
With an average power of 1200 W inside the 
cavity, the count rate observed in the photon detector was $\sim$ 
90 kHz and the background rate of the order of 16 kHz.

For the data presented here,  only the central crystal was
used in the backscattered photon detection.
The electronics was triggered when the amplitude of a
signal recorded in the PMT coupled to that central crystal exceeded 
a preset threshold value.
The state (Cavity ON/Cavity OFF) of a recorded event  is 
determined by the photon beam diagnostics outside the
cavity and by the cavity-locking feedback parameters \cite{nico}. 
The first event selection imposed a minimum current of 5 $\mu$A and 
variations smaller than 3 $\mu$A within each run as well as minor cuts to account for
malfunctions of the experimental setup. 
The modulation of the electron beam position and energy \cite{aniol}
imposed by the HAPPEX experiment induced a significant loss of events ($\sim$ 36 \%)
in limiting systematic effects (cf section \ref{sec:afalse}).

\section{Experimental asymmetry}

\label{sec:afalse}

The raw asymmetry normalized to the 
electron beam intensity, $A_{raw}$,
is determined at 30 Hz for each pair of opposite 
electron helicity states. 
Averaging all pairs for both photon polarizations
yields the mean raw asymmetry (eq. \ref{eq:asyRL})
The mean rates with cavity ON ($r^1$) and
cavity OFF ($r^0$) are obtained by 
averaging over both electron and both photon states 
within each run. The dilution
factor of the asymmetry is thus  
$(1+\frac{B}{S}) = \frac{r^1}{r^1-r^0}$. 
The average signal and background rates were respectively about 1.9 and 0.4
kHz/$\mu$A. The experimental asymmetry of about $0.013$ was determined with a 
relative statistical accuracy $\sim$ 1.4\% within an hour.
Fluctuations of the signal over background ratio $\frac{B}{S}$  
and of the statistical precison of the background asymmetry  estimation $A_B$
within a run account for a small contribution to 
the error budget (cf table \ref{tab:syst}).

Any correlation between the Compton scattering luminosity and 
the electron helicity induces a false asymmetry which  
adds to the physical asymmetry (eq. \ref{eq:aexp}). 
For each pair $i$ of opposite electron helicity, a systematic 
difference of the transverse electron beam position or angle 
$\Delta p_i = \frac{p_i^+ - p_i^-}{2}$ ($p_i=x,y,\theta_x,\theta_y$) 
gives rise to an asymmetry

\begin{equation}
\dis
A_p^i = \left. \frac{1}{r^i (p=p_i)} \frac{\partial r^i}{\partial p} 
\right|_i \Delta p_i, \quad  \text{with} \quad r^i=\frac{N_i}{t_i 
I_i}.
\label{eq:falsasy}
\end{equation}

\begin{figure}
\begin{center}
\epsfig{figure=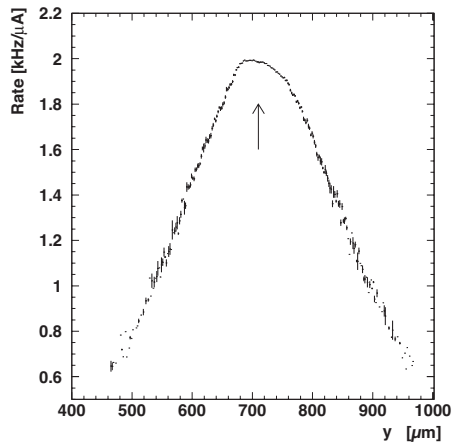,width=0.33\textwidth}
\caption{Count rate normalized to the beam current versus vertical 
position of the electron beam.}
\label{fig:raty}
\end{center}
\end{figure}

The two beams, crossing in the horizontal plane, each have a
transverse size of $\sim$ 100 $\mu$m. The Compton luminosity
is therefore highly sensitive to the vertical
position of the electron beam (fig. \ref{fig:raty}),
of the order of 0.2\%/$\mu$m. Thus, a systematic position
difference of $\sim$ 100 nm would generate a false asymmetry  
of 0.02\%, corresponding to $\sim$ 1.5\% of the experimental asymmetry. 
To minmize the systematic error, the second event selection accepts 
only data lying within 50 $\mu$m from the luminosity optimum.
Again, averaging over the two helicity states of the 
photon beam cancels this contribution if the beam properties 
remain stable over the duration of the run.
However, a slow drift of the beam parameters 
induces a change in the count rate sensitivity. 
The asymmetry induced by a difference $\Delta p_i$ 
can therefore be different for the two photon states, thus creating 
a residual false asymmetry.
This residual contribution is estimated by calculating $A_p^i$
for each pair and summing over the four positions and angles. 
Averaging over all pairs is done using the statistical weights
of the experimental asymmetry.
Helicity-correlated false asymmetries represent on average   
1.2\% of the experimental asymmetry and remain its main 
source of systematic error (cf table \ref{tab:syst}).

\section{Photon beam polarization}

\label{sec:pgamma}

The photon beam polarization can not be measured at the CIP during the 
experiment
since any instrument placed in the photon beam interrupts the build-up 
process.
Instead, the polarization at the exit of the cavity
was monitored on-line during data taking. Absolute polarization measurements 
were performed at the CIP and at the exit of cavity during maintenance 
periods 
when the cavity is removed. These measurements allow  to model 
the transfer function of the light from the CIP to the outside 
detectors, which then result in an evaluation of the photon beam polarization
during the experiment \cite{nico_these}. The polarization
is found to be $\pg^{R,L} = \pm 99.3^{+0.7}_{-1.1}\%$, for both right- 
and left-handed photons.

\input{ac_v2.new.tex}

\section{Results and conclusions}

\label{sec:result}

\begin{figure}
\begin{center}
\epsfig{figure=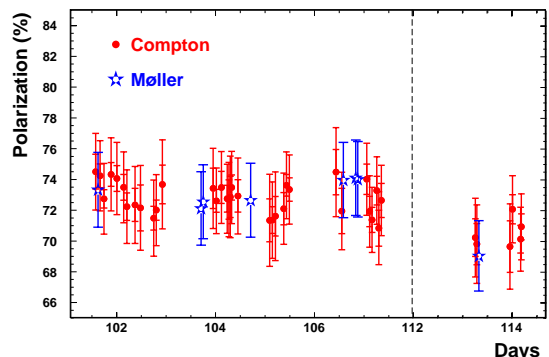,width=0.4\textwidth}
\caption{Polarization  measured by the
Compton (full circles) and the M\o ller (stars) polarimeters during the HAPPEX 
experiment. For the M\o ller data the total error is shown, 
for the Compton the statistical and total error is indicated separately.}
\label{fig:mesure}
\end{center}
\end{figure}

We have measured for the first time the JLab electron beam 
polarization 
at an electron energy of 3.3 GeV and an average current of 40 $\mu$A. 
The 40 measurements of the electron beam polarization,
which was of the order of 70\% \cite{baylac}, are displayed in fig. 
\ref{fig:mesure}.
The average error budget for these data is given in table 
\ref{tab:syst}.
These results are in good agreement with the 
measurements performed with the Hall A M\o ller polarimeter \cite{wwwmoller}.
Moreover, the Compton polarimeter provides a relative monitoring of
the polarization with an error of 2\% due to the
absence of correlations between consecutive runs
in the systematic errors associated with the photon detector ($A_{th}$) and
with the photon polarization ($P_\gamma$).
These results show that the instrument presented here is indeed capable of 
producing a fast and accurate polarization measurement. The
challenging conditions of JLab (100 $\mu$A, few GeV) have
been succesfully adressed with a low-power laser coupled to
a Fabry-Perot cavity. Moreover, the level of systematic 
uncertainties
reached in these first measurements, is encouraging for 
forth-coming achievements. Several hardware improvements have been added 
to the setup since then: new front-end electronic cards,  
feed-back on the electron beam position and a
detector for the scattered electron, consisting of 4 planes of 48 micro-strips. 
Data taken during recent experiments \cite{NDelta}
\cite{GeP} are being analyzed and are expected to
reach a 2\% total error at 4.5 GeV and 100 $\mu A$.

\section{Acknowledgments}

\label{sec:ackno}

The authors would like to acknowledge
Ed Folts and Jack Segal for their constant 
help and support,
as well as the Hall A staff, the HAPPEX 
and Hall A collaboration and the Accelerator Operations group.
This work was supported by the French Commissariat \`{a} l'Energie 
Atomique (CEA) and the Centre National de la Recherche Scientifique 
(CNRS/IN2P3). The Southern Universities Research Association (SURA) 
operates the Thomas Jefferson National Accelerator Facility for the 
US Department of Energy under Contract No. DE-AC05-84ER40150.

\begin{table}
\caption{Average relative error budget.
\label{tab:syst}}
\begin{tabular}{lcccc}
Source & \multicolumn{2}{c}{} & Systematic&Statistical \\
\tableline
$\pg$ & & & 1.1\% & \\
$A_{exp}$ & Statistical    & &  & 1.4\% \\
          & $B/S$          & 0.5\%  & & \\
         & $A_B$          & 0.5\%  & 1.4\% & \\
	  & $\af$          & 1.2\%  & & \\ 
$\ath$     & Non-linearities & 1\% & & \\
          & Calibration           & 1\% & 2.4\% & \\
          & Efficiency/Resolution & 1.9\% &  & \\
\tableline
{\bf Total}  &  &  &  {\bf 3.0}\% &{\bf 1.4}\%   \\
\end{tabular}
\end{table}

\begin{figure}
\begin{center}
\epsfig{figure=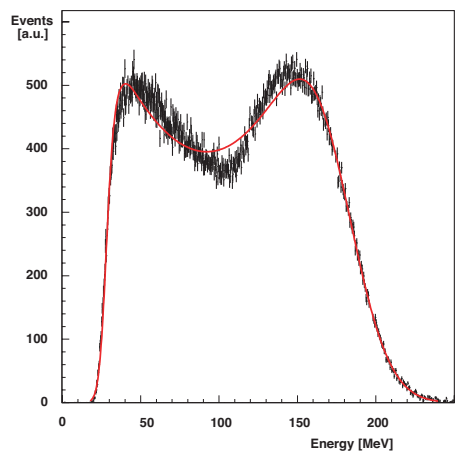,width=0.33\textwidth}
\caption{Energy spectrum of Compton photons, detected in 
the central PbWO$_4$ crystal. The solid curve is the fit, which 
accounts for the calorimeter resolution and efficiency.}
\label{fig:fitcpt}
\end{center}
\end{figure}

\end{document}

%% file: ac_v2.new.tex
\section{Analyzing Power}
The Compton scattering cross section has to be convoluted with the 
response function of the calorimeter:
\[\frac{d\sigma ^{\pm }_{smeared}}{dk_{r}'}=\int ^{\infty }_{o}\frac{d\sigma 
^{\pm }_{c}}{dk'}\, g(k'-k'_{r})\, dk'\]
where \( k' \) is the backscattered photon energy, \( k'_{r} \) is the energy
deposited in the calorimeter, \( \frac{d\sigma ^{\pm }_{c}}{dk'} \) is the
helicity-dependent Compton cross section and \( g(k') \) is the 
response function of the calorimeter. The latter is assumed to be gaussian 
with a width 
\( \sigma _{res}(k')=a\oplus b/\sqrt{k'}\oplus c/k' \)
where \( (a,b,c) \) are fitted to the data. The observed energy spectrum
(fig. \ref{fig:fitcpt}) has a finite width at the threshold. This can be due either to the fact
that the threshold level itself oscillates or to the fact that a given charge
can correspond to different voltage amplitudes at the discriminator level. 
To take this into account the threshold is modelled using an error function 
\( p(k'_{s},k'_{r}) = erf(\frac{k'_{r}-k'_{s}}{\sigma _{s}}) \)
in which \( \sigma _{s} \) is fitted to the data. Hence, the observed count
rate can be expressed as :
\[
N^{\pm }(k'_{s}) = {\cal{L}} \times \int ^{\infty }_{0}p(k'_{s},k'_{r})\, \frac{d\sigma 
^{\pm }_{smeared}}{dk_{r}'}\, dk'_{r}
\]
where $\cal{L}$ is the interaction luminosity. 
An example of a 
measured energy spectrum and a fit using the procedure described in 
this section is shown in fig. \ref{fig:fitcpt}. Finally, the analyzing power of
the polarimeter can be calculated:
\[A_{th}=\frac{N^{+}(k'_{s})-N^{-}(k'_{s})}{N^{+}(k'_{s})+N^{-}(k'_{s})}\]
and is of the order of 1.7 \%. In calculating the energy spectrum, the raw ADC spectrum
had to be corrected for non-linearities in the electronics, resulting 
in a systematic error of 1 \%. Another 1 \% comes from the uncertainty in the calibration
which is performed by fitting the Compton edge. 
Variations of the parameter
\( a,\, b,\, c,\, k'_{s},\, \sigma _{s} \) around the fitted values 
where used to estimate the systematic
error associated with our imperfect modelling of the calorimeter response. 
This contributes  1.9\% to the systematic error which 
is mainly due to the high sensitivity of the analyzing power to the value
of the threshold ($\sim$ 1\%/MeV around 30 MeV). This represents the main
source of systematics in our measurement (cf table \ref{tab:syst}).